\documentclass{article}
\usepackage{graphicx}
\usepackage{dcolumn}
\usepackage{bm}
\usepackage{rotating}
\begin{document}
\begin{center}
    \textbf{The latitude dependence of gravitational redshift from rotating body.}
\end{center}
\begin{center}
    $ Anuj \ Kumar \ Dubey^{\dag}$ and $ Asoke \ Kumar \ Sen^{\S}$
\end{center}
\begin{center}
    \textit{Department of Physics, Assam University, Silchar-788011, Assam, India.}
\end{center}
\begin{center}
    email: $ danuj67@gmail.com^{\dag}$, $ asokesen@yahoo.com^{\S} $
\end{center}
\begin{abstract}
Gravitational redshift is being generally calculated without considering the rotation of a body. Neglecting the rotation, the geometry of space time can be described by using the spherically symmetric Schwarzschild geometry. Rotation has great effect on general relativity, which gives new challenges on gravitational redshift. When rotation is taken into consideration spherical symmetry is lost and off diagonal terms appear in the metric. The geometry of space time can be then described by using the solutions of Kerr family. In the present paper we will discuss the gravitational redshift for rotating body by using Kerr metric. The numerical calculations has been done under Newtonian approximation of angular momentum. The variation of gravitational redshift from equatorial to non - equatorial region has been calculated and its implications are discussed in detail.
\end{abstract}
\section{\label{sec:intro}Introduction}
 General relativity is not only relativistic theory of gravitation proposed by Einstein, but it is the simplest theory that is consistent with experimental data. Predictions of general relativity have been confirmed in all observations and experiments. Gravitational redshift of light is one of the predictions of general relativity and also provides evidence for the validity of the principle of equivalence. Any relativistic theory of gravitation consistent with the principle of equivalence will predict a redshift. \\ If, however, we observe on the Earth the spectrum emitted by the atoms located on the Sun, then, its lines appear to be shifted with respect to the lines of the same spectrum emitted on the Earth. Each line with frequency $\omega$ will be shifted through the interval $\Delta\omega$ given by formula,
 \begin{equation}
 \Delta\omega=\frac{\omega(V_{1}-V_{2})}{c^{2}}
 \end{equation}
 where $V_{1}$ and $V_{2}$ are the potentials of the gravitational field at the points of emission and observation of the spectrum respectively (page 269, of Landau and Lifshitz [1]).
 If we observe on the earth a spectrum emitted on the sun or the stars, then magnitude of $V_{1}$ is greater than magnitude of $V_{2}$  and it is clear that $\Delta\omega<$0, i.e, the shift occurs in the direction of lower frequency. The phenomenon we have described is called the \lq gravitational  redshift\rq. \\ Adams in 1925 has claimed first about the confirmation of the predicted gravitational redshift from the measurement of the apparent radial velocity of Sirius B [2].
 Pound and Rebka in 1959 were the first to experimentally verify the gravitational redshift from nuclear - resonance [3]. Pound and Snider in 1965 had performed an improved version of the experiment of Pound and Rebka, to measure the effect of gravity, making  use of Mossbauer - Effect [4].
  If the Source of radiation is at a height \textit{h} above an observer then the result found was (0.9990 $\pm$ 0.0076) times the value of $4.905\times 10^{-15}\times\frac{2gh}{c^{2}}$ predicted from principle of equivalence. Snider in 1972 has measured the redshift of the solar potassium absorption line at 7699 ${\AA}$ by using an atomic - beam resonance - scattering technique [5].
  Krisher et al. in 1993 had measured the gravitational redshift of Sun [6].\\ Gravitational redshift has been reported by most of the authors without consideration of rotation of a body. Neglecting the rotation, the geometry of space time can be described using the well-known spherically symmetric Schwarzschild's geometry and information on the ratio $ \frac{M}{r}$ of a compact object can be obtained from the gravitational redshift, where M and r are mass and radius respectively. P D Nunez and M Nowakowski in 2010  had obtained an expression for gravitational redshift factor of rotating body by using small perturbations to the Schwarzschild's geometry. Two main results of this work include the derivation of a maximum angular velocity depending only on the mass of the object and a possible estimates of the radius [7]. \\
  With this background the present paper is organized as follows. In Section - 2, we  derive the expression for gravitational redshift. In Section - 3, the derived expression of gravitational redshift is applied to calculate the values of gravitational redshift for  various rotational bodies which includes the Sun and many millisecond pulsars. In Section - 4, we calculate the red-shifted wavelength ratio, if photon is emitted from equatorial and polar regions. We further discuss its variation with rotation parameter for different values of latitude at which photon is emitted. Finally, some discussions and conclusions are made in Section - 5.
  \section{\label{sec:Grav redshift} Gravitational redshift from rotating body}
\subsection{ Kerr Field}
When rotation is taken into consideration spherical symmetry is lost and off - diagonal terms appear in the metric and the most useful form of the solution of Kerr family is given in terms of t, r, $\theta$ and $\phi$, where t, and r are Boyer - Lindquist coordinates running from - $\infty$ to + $\infty$, $\theta$  and $\phi$, are ordinary spherical coordinates in which  $\phi$ is periodic with period of 2 $\pi $ and  $\theta$ runs from 0 to  $\pi$.
\\Covariant form of metric tensor with signature (+,-,-,-) is expressed as
\begin{equation}
ds^{2} = g_{tt}c^{2}dt^{2}+ g_{rr} dr^{2} + g_{\theta\theta} d\theta^{2} +g_{\phi\phi} d\phi^{2} + 2 g_{t \phi} c dt d\phi
\end{equation}
\\ Non-zero components $g_{ij} $ of Kerr family are given as follows (page 346, of Landau and Lifshitz [1] and page 261-263, of Carroll [8]),
\begin{equation}
 g_{tt}=\frac{\Delta-a^{2}sin^{2}\theta}{\rho^{2}} = 1-\frac{r_{g} r}{\rho^{2}}
\end{equation}
\begin{equation}
g_{rr}=-\frac{\rho^{2}}{\Delta}
\end{equation}
\begin{equation}
g_{\theta\theta}=-{\rho^{2}}
 \end{equation}
 \begin{equation}
g_{\phi\phi}=-\frac{[(r^{2}+a^{2})^{2}-a^{2}\Delta sin^{2}\theta]sin^{2}\theta}{\rho^{2}}
 \end{equation}
 \begin{equation}
g_{t\phi}=\frac{a sin^{2}\theta (2mr-e^{2})}{\rho^{2}}
\end{equation}
with
\begin{equation}
\rho^{2} = r^{2}+ a^{2}cos^{2}\theta
\end{equation}
\begin{equation}
 \Delta = r^{2}-2mr + a^{2}+e^{2} =r^{2}-r_{g}r + a^{2}+e^{2}
\end{equation}
where the parameters $ 2m (=\frac{2GM}{c^{2}})$, e, and $ a (= \frac{J}{Mc})$ are respectively Scharzschild radius ($r_{g}$), charge and rotation parameter of the source.  J is the angular momentum of the central body, which can be also written as $J = I\Omega$. I and $\Omega$ are the moment of inertia and angular velocity respectively. Charge (e) is written as $ e = \sqrt {Q^{2}+P^{2}} $ , where Q is electric charge and P is magnetic charge. In case of Kerr solution in 1963 both Q and P vanishes [9], for Kerr - Newman solution in 1965 only P vanish [10] but in case of Kerr - Newman - Kasuya solution in 1982 both Q and P are non-vanishing [11].\\
\subsection{Four - Velocity in Kerr Field}
Four - dimensional velocity (four - velocity) of a particle is defined as the four - vector (page 23, of Landau and Lifshitz [1])
\begin{equation}
u^{i} =\frac{dx^{i}}{ds}
\end{equation}
 where the index i takes on the values 0,1,2,3 and $x^{0} = ct$, $x^{1} = r$, $x^{2} = \theta$, $x^{3} = \phi$.
 \\ Using equation (2) we can write
 \begin{equation}
ds =\sqrt{g_{tt}c^{2}dt^{2}+ g_{rr} dr^{2} + g_{\theta\theta} d\theta^{2} +g_{\phi\phi} d\phi^{2} + 2 g_{t \phi} c dt d\phi}
\end{equation}
The components of four - velocity are
\begin{equation}
 u^{0}\equiv u^{t} =\frac{dx^{o}}{ds} = \frac{cdt}{\sqrt{g_{tt}c^{2}dt^{2}+ g_{rr} dr^{2} + g_{\theta\theta} d\theta^{2} +g_{\phi\phi} d\phi^{2} + 2 g_{t \phi} c dt d\phi}}
\end{equation}
\begin{equation}
 u^{1}\equiv u^{r} =\frac{dx^{1}}{ds}  = \frac{dr}{\sqrt{g_{tt}c^{2}dt^{2}+ g_{rr} dr^{2} + g_{\theta\theta} d\theta^{2} +g_{\phi\phi} d\phi^{2} + 2 g_{t \phi} c dt d\phi}}
\end{equation}
\begin{equation}
 u^{2}\equiv u^{\theta} =\frac{dx^{2}}{ds} = \frac{d\theta}{\sqrt{g_{tt}c^{2}dt^{2}+ g_{rr} dr^{2} + g_{\theta\theta} d\theta^{2} +g_{\phi\phi} d\phi^{2} + 2 g_{t \phi} c dt d\phi}}
\end{equation}
\begin{equation}
 u^{3}\equiv u^{\phi} =\frac{dx^{3}}{ds} = \frac{d\phi}{\sqrt{g_{tt}c^{2}dt^{2}+ g_{rr} dr^{2} + g_{\theta\theta} d\theta^{2} +g_{\phi\phi} d\phi^{2} + 2 g_{t \phi} c dt d\phi}}
\end{equation}
From equations (12) and (15) we can write
\begin{equation}
\frac{u^{3}}{u^{0}}\equiv\frac{u^{\phi}}{u^{t}} = \frac{d\phi}{c dt} = \frac{\Omega}{c}
\end{equation}
 where\begin{equation}
\Omega = \frac{d\phi}{dt}\end{equation}
For a sphere, the photon is emitted at a location on its surface where $dr = d\theta =0$, when the sphere rotates.
\\Now using equation (16) the expressions for four - velocity becomes,
 \begin{equation}
 u^{0}\equiv u^{t} =\frac{1}{\sqrt{g_{tt}+ g_{\phi\phi}\frac{\Omega^{2}}{c^{2}} +2 g_{t \phi}\frac{\Omega}{c}}}
\end{equation}
\begin{equation}
u^{1}\equiv u^{r}=0
\end{equation}
\begin{equation}
u^{2}\equiv u^{\theta} = 0
\end{equation}
\begin{equation}
 u^{3}\equiv u^{\phi} = u^{t} \frac{\Omega}{c}
\end{equation}
Hence four - velocity of an object in Kerr field can be expressed as,
\begin{equation}
u^{i} = (u^{t},0,0,u^{t} \frac{\Omega}{c})
\end{equation}
\subsection{Frequency in Kerr Field}
Let f be any quantity describing the field of the wave. For a plane monochromatic wave f has the form (page 140, of Landau and Lifshitz [1])
\begin{equation}
f= a e^{\textbf{j}(k.r-\omega t +\alpha)} =  a e^{\textbf{j}(k_{i}x^{i} +\alpha)}
\end{equation}
where $\textbf{j} = \sqrt{-1}$, and i used as superscript and subscript indicates i=0,1,2,3. \\
k is propagation constant and $k_{i}$ is the wave four - vector.
\\We write the expression for the field in the form
\begin{equation}
f = a e^{\textbf{j}\Psi}
 \end{equation}
where $\Psi\equiv -k_{i}x^{i} +\alpha$, is defined as eikonal.
\\Over small space region and time intervals the eikonal $\Psi$ can be expanded in series; to terms of first order, we have
\begin{equation}
\Psi=\Psi_{0}+r.\frac{\partial\Psi}{\partial r}+t\frac{\partial\Psi}{\partial t}
\end{equation}
As a result one can write (page 141, of Landau and Lifshitz [1])
\begin{equation}
k_{i}=-\frac{\partial\Psi}{\partial x^{i}}
\end{equation}
where $k_{i}$ is the wave four - vector and the components of the four - wave vector $k_{i}$ are related by
\begin{equation}
{k_{i}}{k^{i}} = 0
\end{equation}
or
\begin{equation}
\frac{\partial\Psi}{\partial x^{i}}\frac{\partial\Psi}{\partial x_{i}}=0
\end{equation}
This equation called the eikonal equation is fundamental equation of geometrical optics.\\
 Using equation (26), the components of wave four - vector are
\begin{equation}
k_{0}=-\frac{\partial\Psi}{\partial x^{0}}= -\frac{\partial\Psi}{\partial ct}=\frac{\omega}{c}
\end{equation}
\begin{equation}
k_{1}=k_{x}= -\frac{\partial\Psi}{\partial x^{1}}= -\frac{\partial\Psi}{\partial x}
\end{equation}
\begin{equation}
k_{2}=k_{y}= -\frac{\partial\Psi}{\partial x^{2}}= -\frac{\partial\Psi}{\partial y}
\end{equation}
\begin{equation}
k_{3}=k_{z}= -\frac{\partial\Psi}{\partial x^{3}}= -\frac{\partial\Psi}{\partial z}
\end{equation}
Hence we can write the wave - four vector
 \begin{equation}
k_{i}= (k_{0},k_{1},k_{2},k_{3})=( \frac{\omega}{c},k_{x},k_{y},k_{z})
\end{equation}
$k_{0}(=\frac{\omega}{c})$ is the time component of four - wave vector and this frequency is measured in terms of the world time.
\\Frequency measured in terms of proper time ($\tau$) is defined as
\begin{equation}
\omega^{'}=-\frac{\partial\Psi}{\partial \tau}
\end{equation}
This frequency $ \omega^{'} $ is different at different point of space.
\begin{equation}
\omega^{'}= -\frac{\partial\Psi}{\partial x^{0}}\frac{\partial x^{0}}{\partial \tau}  = \omega u^{0}
\end{equation}
as $\frac{\partial x^{0}}{\partial \tau}$ is nothing but $u^{0}$.\\
Substituting the value of $ u^{0} $ in above equation from equation (18)
\begin{equation}
\omega^{'}=\frac{\omega}{\sqrt{g_{tt}+ g_{\phi\phi}\frac{\Omega^{2}}{c^{2}} +2 g_{t \phi}\frac{\Omega}{c}}}
\end{equation}
This $\omega^{'}$ is the frequency measured by a distant observer with proper time ($\tau$). Where $\omega$ is the frequency measured in terms of world time $t(=\frac{x^{0}}{c}$).\\
\\\textbf{ Alternative method: Frequency in Kerr field}\\
The Lagrangian $\pounds$ of the test particle can be expressed as
\begin{equation}
\pounds  = \frac{g_{ij}\dot{x}^{i}\dot{x}^{j}}{2}
\end{equation}
(Dot over a symbol denotes ordinary differentiation with respect to an affine parameter $\xi$)\\
From Lagrangian of the test particle, we can obtain the momentum of the test particle as,
\begin{equation}
P_{i} = \frac{\partial\pounds}{\partial\dot{x}^{i}}= g_{ij} \dot{x}^{j}
\end{equation}
Thus corresponding momentum in the coordinates of t, r, $\theta$, and $\phi$ are given by
\begin{equation}
P_{t} \equiv - E= g_{tt} \dot{t} + g_{t\phi} \dot{\phi}
\end{equation}
\begin{equation}
P_{r} = g_{rr} \dot{r}
\end{equation}
\begin{equation}
P_{\theta} = g_{\theta\theta} \dot{\theta}
\end{equation}
\begin{equation}
P_{\phi} \equiv L= g_{t\phi} \dot{t} + g_{\phi\phi} \dot{\phi}
\end{equation}
Since the space time of the Kerr family is stationary and axially symmetric, the momenta $P_{t}$ and $P_{\phi}$  are conserved along the geodesics. So we obtain two constants of motion: one is corresponding to the conservation of energy (E) and the other is the angular momentum (L) about the symmetry axis.\\
Any observer measures the frequency ($\omega^{'}$) of a photon following null geodesic $x^{i} (\xi)$ can be calculated by the expression given as (page 217 of Carroll 2004 [8] and page 108 of Straumann 1984 [12],
\begin{equation}
\omega^{'} = u^{i}\frac{dx_{i}}{d\xi}=u^{i} g_{ij}\frac{dx^{j}}{d\xi}
\end{equation}
\begin{equation}
\omega^{'} = u^{t}(g_{tt} \dot{t} + g_{t\phi} \dot{\phi})+u^{\phi}(g_{t\phi} \dot{t} + g_{\phi\phi} \dot{\phi})
\end{equation}
using equation (39) and (42) in above equation (44), we can write
\begin{equation}
\omega^{'} = u^{t}(-E) + u^{\phi}(L)
\end{equation}
using equation (16) in above equation (45)
\begin{equation}
\omega^{'} = u^{t}(-E + \frac{\Omega}{c} L)
\end{equation}
\begin{equation}
\omega^{'} = \omega u^{0}
\end{equation}
Here we define $(-E + \frac{\Omega}{c}L)$ as $\omega$, where $\omega$ can be identified as the frequency corresponding to world time as in equation (29) and equation (35).
\\ As a result, we can write the expression of frequency $(\omega^{'})$ observed by a distant observer as,
\begin{equation}
\omega^{'}=\frac{\omega}{\sqrt{g_{tt}+ g_{\phi\phi}\frac{\Omega^{2}}{c^{2}} +2 g_{t \phi}\frac{\Omega}{c}}}
\end{equation}
From the given two different approaches the expression of frequency (36) and (48) look exactly similar.
\\In General relativity redshift (Z) and redshift factor ($\Re$) are defined as [13],
\begin{equation}
\frac{1}{Z+1}= \Re =\frac{\omega_{ob}}{\omega_{em}}=\frac{\omega^{'}}{\omega} = \frac{\lambda_{em}}{\lambda_{ob}}
\end{equation}
where $\omega$ and $\lambda$ denote frequency and wavelength, respectively, and the redshift and redshift factor are Z and $\Re$ respectively. Emitter's and observer's frame of reference are indicated by subscripts \textit{em} (unprimed $\omega$) and \textit{ob} (primed $\omega$). A redshift of zero corresponds to an un-shifted line, whereas $Z<0$ indicates blue-shifted emission and $Z>0$ red-shifted emission. A redshift factor of unity corresponds to an un-shifted line, whereas $\Re<1$ indicates red-shifted emission and $\Re>1$ blue-shifted emission.
\\ From above equations (48) and (49), we can write redshift factor ($\Re$),
\begin{equation}
\Re= \sqrt{g_{tt}+ g_{\phi\phi}\frac{\Omega^{2}}{c^{2}} +2 g_{t \phi}\frac{\Omega}{c}}
\end{equation}
Using equation (49), we can obtain redshift (Z) from above redshift factor ($\Re$) equation (50).
\section{\label{sec:Num cal Grav redshift} Numerical Calculation of Gravitational redshift from rotating bodies}
If we ignore the electric and magnetic charges contribution and then using the metric coefficients $g_{tt}$, $g_{\phi\phi}$ and $g_{t \phi}$ of Kerr family in equation (50), we can obtain the redshift factor ($\Re$) and corresponding redshift (Z) for the values of $\theta$ (in degree) from zero (pole) to $\frac{\pi}{2}$ (equator).
\\For doing numerical calculations we have considered the Newtonian approximation of angular momentum $J (= \frac{2Mr^{2}\Omega}{5})$.
\\So rotation parameter can be written as
 \begin{equation}
 a=\frac{2r^{2}\Omega}{5c}
 \end{equation}
 where $\Omega$ is angular velocity as described in Section III A.
 \\Observed or redshifted Lyman - $\alpha$ line (emitted wavelength of 1215.668 ${\AA}$) can be calculated from equation (49) as,
\begin{equation}
\lambda_{ob} =\frac{\lambda_{em}}{\Re} = \lambda_{em} (Z+1)
\end{equation}
\begin{table*}
\caption{\label{tab:table1}A list showing  Mass (M), Schwarzschild radius ($r_{g}$), Radius (r), Angular velocity ($\Omega$), and Rotation parameter (a) for different millisecond Pulsars and Sun. Data taken from various authors as cited in column 2.}
\begin{tabular}{|c|c|c|c|c|c|c|}
                      \hline
                      S. No. & Star  & M$(M_{Sun})$ & $r_{g}$(km) & r(km) & $\Omega$(rad/s) & a(km) \\ \hline
                      1 & PSR J 1748-2446ad [14] & 1.350 & 4.050 & 20.1 & 4.4985$\times10^{3}$  & 2.42325   \\
                      2 &PSR B 1937+21 [15]& 1.350 & 4.050 & 20.2 & 4.0334$\times10^{3}$  & 2.19438  \\
                      3 & PSR J 1909-3744 [16] & 1.438 & 4.314 & 31.1 & 2.1300$\times10^{3}$  & 2.74687  \\
                      4 & PSR 1855+09 & 1.350 & 4.050 & 46.9 & 1.1849$\times10^{3}$  & 3.47509   \\
                      5 & PSR J 0737-3039 A [17] & 1.340 & 4.020 & 133.6 & 0.2766$\times10^{3}$ & 6.58269   \\
                      6 & PSR 0531+21 & 1.350 & 4.050 & 164.4 & 0.1885$\times10^{3}$  & 6.79287   \\
                      7 & PSR B 1534+12 [18] & 1.340 & 4.020 & 165.9 & 0.1657$\times10^{3}$  & 6.08070   \\
                      8 & PSR B 1913+16 & 1.440 & 4.320 & 10.0 & 0.1064$\times10^{3}$  & 0.01418  \\
                      9 & Sun & 1.000 & 3.000 & 0.7$\times10^{6}$ & 2.5982$\times10^{-6}$  & 1.69749 \\ \hline
\end{tabular}
\end{table*}
\begin{sidewaystable}
\caption{\label{tab:table2}Calcualted values of  Redshift factor ($\Re$), Redshift (Z) and corresponding redshifted  wavelength of Lyman - $\alpha$ line (emitted wavelength of 1215.668 ${\AA}$) for Sun.}
\begin{tabular}{|c|c|c|c|c|c|c|}
                      \hline
                       $\Re(\theta =0$) & $\Re(\theta =\frac{\pi}{2}$) & Z($\theta =0$) & Z($\theta =\frac{\pi}{2}$) & $\kappa(\theta=0)$&  $ Ly\ - \alpha, \ \lambda (\theta =0){\AA}$ & $  Ly\ - \alpha, \ \lambda (\theta =\frac{\pi}{2}){\AA}$\\\hline
                       0.9999978825374& 0.9999978825220 & 2.117467$\times10^{-6}$ & 2.117482$\times10^{-6}$ & 1.000000000015 & 1215.67054288& 1215.67054290 \\                       \hline
\end{tabular}
\end{sidewaystable}
\begin{figure}
\includegraphics[width=30pc, height=30pc,angle=270]{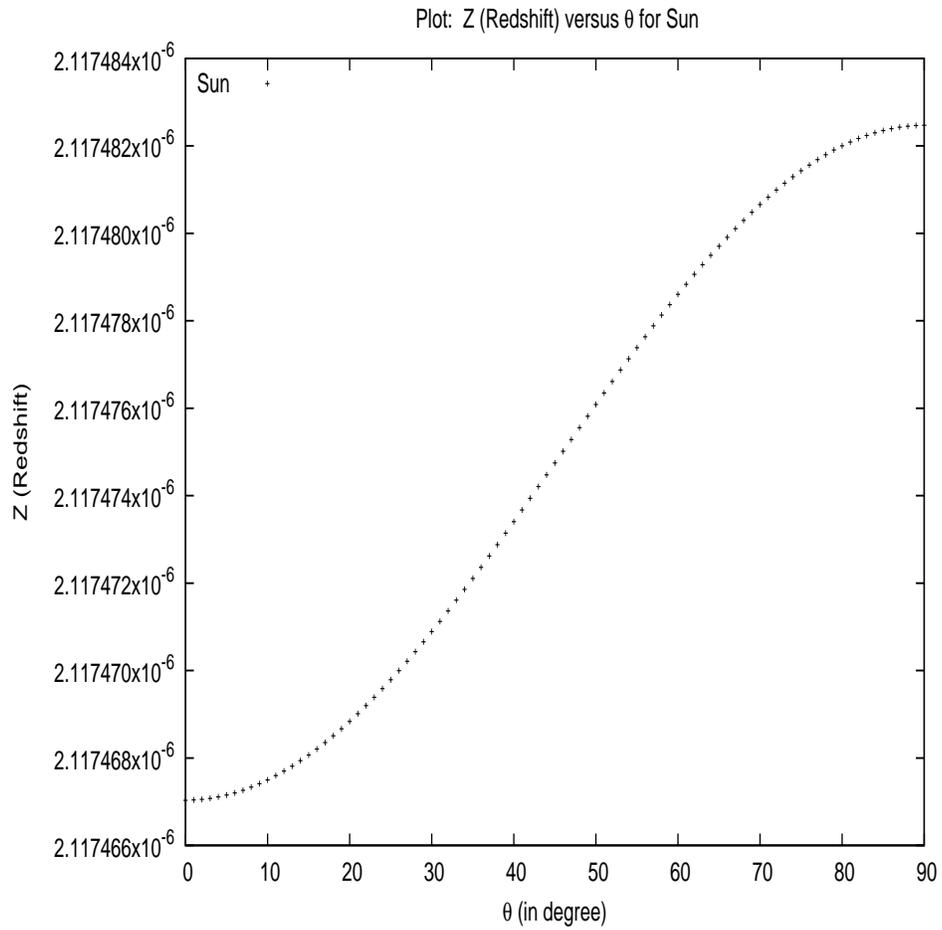}
\caption{\label{fig1} Shows the variation of redshift ($Z$) for Sun from polar to equatorial region, where the ($90^o$ - latitude) values or $\theta$ in degree varies  from 0 to $\frac{\pi}{2}$ .}
\end{figure}
\begin{sidewaystable}
\caption{\label{tab:table3}Calculated values of  Redshift factor ($\Re$), Redshift (Z) and corresponding redshifted  wavelength of Lyman - $\alpha$ line (emitted wavelength of 1215.668 ${\AA}$) for different millisecond Pulsars.}
\begin{tabular}{|c|c|c|c|c|c|c||c||c|}
                      \hline
                      S. No. & Star  & $\Re(\theta =0$) & $\Re(\theta =\frac{\pi}{2}$) & Z ($\theta =0$) & Z ($\theta =\frac{\pi}{2}$) & $\kappa(\theta=0)$ &  $ Ly\ -  \alpha, \ \lambda  (\theta =0){\AA}$ & $  Ly\ -\alpha, \ \lambda (\theta =\frac{\pi}{2}){\AA}$ \\ \hline
                      1 & PSR J 1748-2446ad  & 0.896519& 0.858760 & 0.115424 & 0.164469  & 1.0439701&1355.9854& 1415.6083   \\
                      2 &PSR B 1937+21& 0.896767 & 0.866352& 0.115116 & 0.154264 &1.0351065& 1355.6111&1403.2019 \\
                      3 & PSR J 1909-3744 & 0.929511& 0.909332& 0.075834 & 0.099707 & 1.0221908& 1307.8571& 1336.8796 \\
                      4 & PSR 1855+09 & 0.956627 &0.942328& 0.045339 & 0.061201 & 1.0151740 & 1270.7856& 1290.0685 \\
                      5 & PSR J 0737-3039 A &0.985057 & 0.978695 & 0.015168 & 0.021768& 1.0065012& 1234.1083& 1242.1315  \\
                      6 & PSR 0531+21 & 0.987774& 0.983301 & 0.012376 & 0.016981 & 1.0045485 & 1230.7140& 1233.9878 \\
                      7 & PSR B 1534+12 & 0.987971 & 0.984455& 0.012174 & 0.015790 & 1.0035720&1230.4685& 1234.8638 \\
                      8 & PSR B 1913+16 &0.757047   & 0.757042& 0.320921  & 0.320929 & 1.0000063& 1605.8015& 1605.8117 \\
                       \hline
\end{tabular}
\end{sidewaystable}
\begin{figure}
\includegraphics[width=30pc, height=30pc,angle=270]{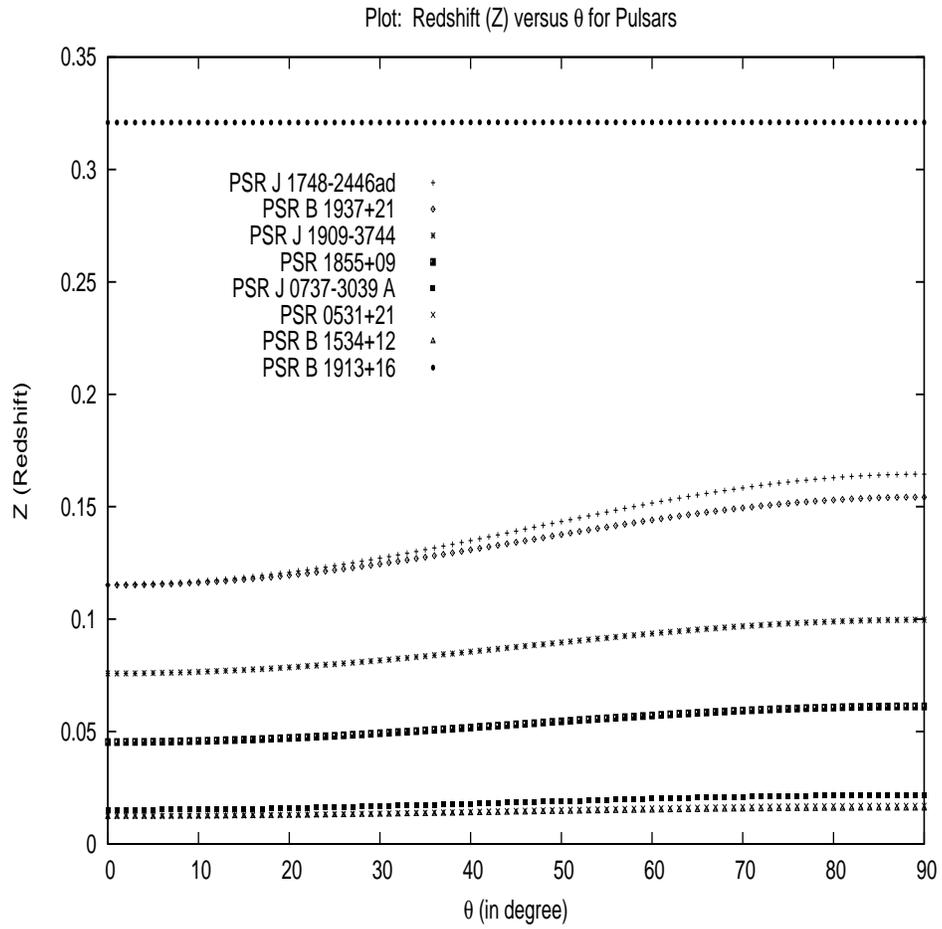}
 \caption{\label{fig2} Shows the variation of redshift for different millisecond Pulsars from equatorial to polar region.}
\end{figure}
\section{\label{sec:Redshifted wavelength} Redshifted wavelength ratio at equator and pole}
From equation (49), we can write
\begin{equation}
\frac{\lambda_{ob, \theta=90}}{\lambda_{ob, \theta}} =\frac{\Re_{ob, \theta}}{\Re_{ob, \theta=90}}=\kappa(\theta) (say)
\end{equation}
The quantity $\kappa(\theta)$ can be termed as \lq coefficient of latitude  dependence of redshift for rotating mass\rq. \\
Using the value of $\Re$ from equation (50) in above equation (53), we can write
\begin{equation}
 \kappa(\theta) = \frac{[\sqrt{g_{tt}+ g_{\phi\phi}\frac{\Omega^{2}}{c^{2}} +2 g_{t \phi}\frac{\Omega}{c}}] _ {\theta}}{[\sqrt{g_{tt}+ g_{\phi\phi}\frac{\Omega^{2}}{c^{2}} +2 g_{t \phi}\frac{\Omega}{c}}]_ {\theta=\frac{\pi}{2}}}
 \end{equation}
 We can find $\kappa(\theta)$ for $\theta =0$, which will give a ratio of observed redshifted wavelength between polar and equatorial region.\\
  The values of $g_{tt}$, $g_{t\phi}$ and $g_{\phi\phi}$ for polar and equatorial regions can be obtained from equations (3) to (9).
 \\ For equatorial plane where $\theta =\frac{\pi}{2} $, from equations (8) and (9), if we consider charge e = 0, then $ \rho^{2} = r^{2}$  and $  \Delta = r^{2}-2mr + a^{2} =r^{2}-r_{g}r + a^{2} $.
 \\ From equations (3), (6) and (7) metric elements $ g_{tt}$, $ g_{\phi\phi}$ and $g_{t\phi}$ for Kerr space time at equator becomes,
 \begin{equation}
 g_{tt}(\theta =\frac{\pi}{2})=1-2\frac{m}{r} = 1-\frac{r_{g}}{r}
\end{equation}
\begin{equation}
 g_{\phi\phi}(\theta =\frac{\pi}{2})= - (r^{2}+a^{2}+\frac{2ma^{2}}{r}) = -(r^{2}+a^{2}+\frac{r_{g}a^{2}}{r})
\end{equation}
\begin{equation}
g_{t\phi}(\theta =\frac{\pi}{2})= \frac{2ma}{r} =\frac{r_{g}a}{r}
\end{equation}
\\ At poles where $\theta =0 $, from equations (8) and (9), if we consider charge e = 0, then $ \rho^{2} = r^{2}+a^{2}$  and $  \Delta = r^{2}-2mr + a^{2} =r^{2}-r_{g}r + a^{2} $.
\\ From equations (3), (6) and (7) metric elements $ g_{tt}$, $ g_{\phi\phi}$ and $g_{t\phi}$ for Kerr space time at poles becomes,
\begin{equation}
 g_{tt}(\theta=0)=1-2\frac{mr}{r^{2}+a^{2}} = 1-\frac{r_{g}r}{r^{2}+a^{2}}
\end{equation}
\begin{equation}
g_{\phi\phi}(\theta=0)= 0
\end{equation}
\begin{equation}
g_{t\phi}(\theta=0)= 0
\end{equation}
Substituting the values of $g_{tt}$, $g_{t\phi}$ and $g_{\phi\phi}$ from equations (55) to (60) in equation (54), we get
\begin{equation}
\kappa(\theta=0) =\frac{\sqrt{(1-\frac{r_{g}r}{r^{2}+a^{2}}})}{\sqrt{(1-\frac{r_{g}}{r})+\frac{5r_{g}a^{2}}{r^{3}} -[{r^{2}+a^{2}(1+\frac{r_{g}}{r})}]{\frac{25a^{2}}{4r^{4}}}}}
\end{equation}
The redshifted values that we often refer for rotating mass is actually that corresponding to $\theta =\frac{\pi}{2}$. Whereas at $\theta =0$, it will represent a value corresponding to Schwarzschild mass.
\\ The difference between the redshifts as observed from pole and equator can be normalized by dividing  it by the redshift as observed at equator. Such a parameter if we denote by $\eta$, then we can write
\begin{equation}
\eta = \kappa(\theta=0)-1
\end{equation}
We can make a plot of $\eta$ v/s a (rotation parameter) by  keeping $r_{g}$ fixed for Sun, which is $\sim$ 3 km. The plot has been made for values of $r= 10^{2}r_{g}$ onwards. However, as the radius of Sun is $0.7\times10^{6}$ km, so values of r, greater than or equal to $\sim 2.3\times10^{5}r_{g}$ are only physically meaningful. For smaller value of r, these plots are for research interest. However, for other compact objects (such Pulsar) having similar $r_{g}$ and a (rotation parameter) values as that of Sun, there plots will be useful. Cases of few such Pulsars are considered in next section.
\begin{figure}
\includegraphics[width=30pc, height=30pc,angle=270]{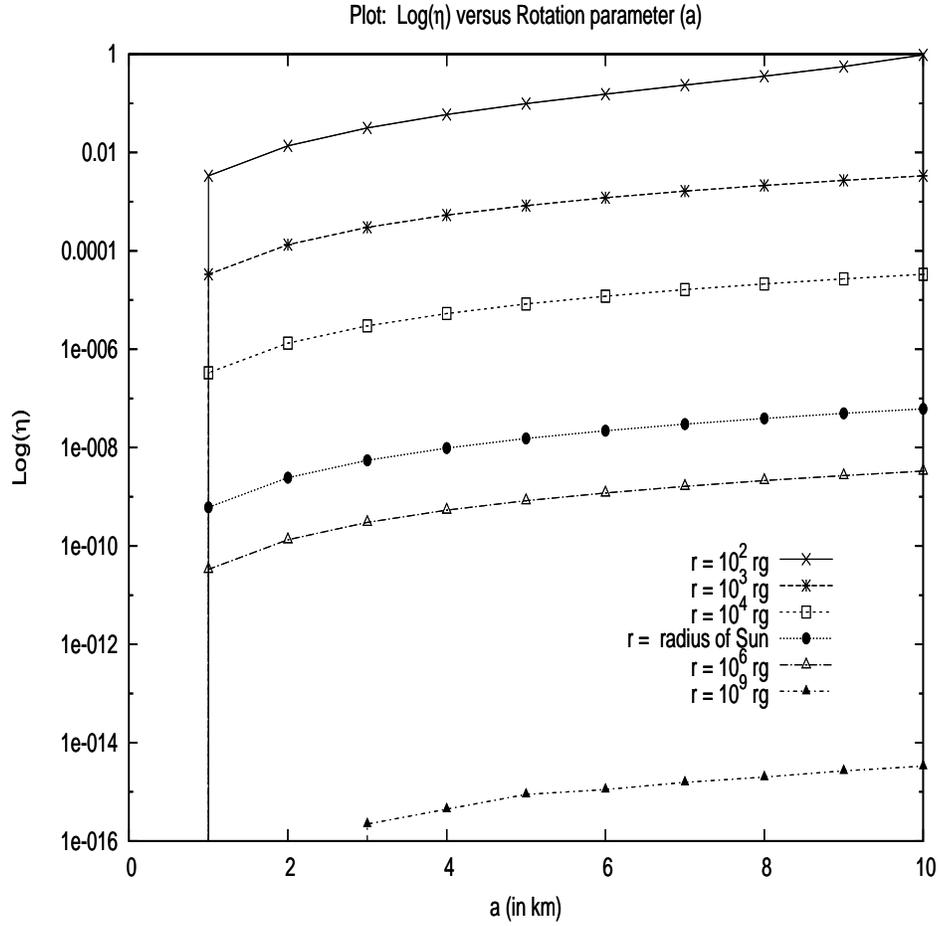}
 \caption{\label{fig3} Shows plot of Log($\eta$) versus the rotation parameter (a). The  value of  $r_{g}$ has been kept fixed at that of Sun=3 km. The different curves are for different values of $r= 10^{2}r_{g}$,$ 10^{3}r_{g}$,$ 10^{4}r_{g}$,$ 10^{6}r_{g}$,$ 10^{9}r_{g}$ and for the actual radius of Sun.}
\end{figure}
\section{\label{sec:Concl}Conclusions and Discussions}
 We can conclude from equation (50) derived in the present work that, the gravitational redshift is affected by rotation of source because the redshift factor expression contains the angular velocity of rotation of source ($\Omega $), which is in powers of two and four. As a result the observed redshift will not depend on the direction of rotation of the source. Further one can see from Eqns. (49) and (50) that, the redshit is a function of the latitude ($90^o - \theta$). When we consider the angular velocity of rotation of source to be zero  ($\Omega =0$), then we can obtain the corresponding gravitational redshift from a static body of same mass (Schwarzschild Mass). The same expression is also obtained if we consider a Kerr mass , but with the boundary condition $\theta$  equal to zero (polar region). Gravitational redshift factor equation (50) also indicates that gravitational redshift depends on electric and magnetic charges contribution due to presence of electric and magnetic charges in $g_{tt}$, $g_{t\phi}$ and $g_{\phi\phi}$.
 \\A set of eight Pulsars for which different parameters like rotation and radius values are available in literature have been considered for some numerical calculations, in the present work. The idea is to apply the calculations made in the present work to the cases of such Pulsars. The calculated numerical values of gravitational redshift (Z) for Sun and these millisecond Pulsars (PSR J 1748-2446ad, PSR B 1937+21, PSR J 1909-3744, PSR 1855+09, PSR J 0737-3039 A, PSR 0531+21, PSR B 1534+12 and PSR B 1913+16) are  given in TABLE - II and TABLE - III. The  corresponding plots of $\theta$ versus Z is shown in FIG - 1 and FIG - 2. From TABLE - II, TABLE - III, FIG - 1 and FIG - 2, it is clearly seen that the gravitational redshift is highest at equator and lowest at pole. The plot of rotation parameter (a) versus Log($\eta$), is shown in FIG - 3. From FIG - 3, we can conclude that with higher the rotation parameter (a), $\eta$ also becomes higher.
 \\From TABLE - II, TABLE - III, it is also seen that redshifted or observed wavelength of  Lyman - $\alpha$ line (emitted wavelength of 1215.668 ${\AA}$) for Sun and many millisecond Pulsars is maximum at equator and minimum at poles.
 \\The differences between the redshifted Lyman - $\alpha$ wavelength (between pole and equator) for most of these  Pulsars are a few Angstroms or few tens of Angstroms. However, for Sun as seen from table II, this difference comes only at the seventh place after the decimal point in Angstrom, thus it would remain undetectable to any modern spectrograph. However, for these eight Pulsars PSR J 1748-2446ad, PSR B 1937+21, PSR J 1909-3744, PSR 1855+09, PSR J 0737-3039 A, PSR 0531+21, PSR B 1534+12 and PSR B 1913+16, the differences are 59.6229, 47.5908, 29.0224, 19.2829, 8.0232, 3.2737, 4.3952 and 0.0102 in Angstrom respectively.  As a result with modern spectrometers (as back end instruments to many telescopes), it will be possible to distinguish between the redshifted Lyman - $\alpha$ emitted from the polar and equatorial region, subject to the condition that the Pulsar is emitting in Lyman - $\alpha$ in the optical. Thus the calculations made in this paper will help us to identify the region of line emission from  various rotating objects. This will also in turn verify the calculations made in the present work.
\section{\label{sec:ack}Acknowledgments}
We would like to thank  Atri Deshmukhya, Head  Department of Physics, Assam University, Silchar, India for useful discussions and suggestions. A Dubey is also thankful to Sanjib Deb and Arindwam Chakraborty, Department of Physics, Assam University, for providing help and support in programming and plots.

\end{document}